\documentclass[10pt,letterpaper,twocolumn]{article} 

\usepackage{ol2}
\usepackage[draft,implicit=false]{hyperref}
\usepackage{amsmath}

\begin{document}

\twocolumn[ 

\title{Tunable terahertz coherent perfect absorption in a monolayer graphene}

\author{Yuancheng Fan,$^{1,*}$ Fuli Zhang,$^{1}$ Qian Zhao,$^{2}$ Zeyong Wei,$^{3}$ and Hongqiang Li$^{3}$}

\address{
$^1$Key Laboratory of Space Applied Physics and Chemistry, Ministry of Education and\\ Department of Applied Physics, School of Science, Northwestern Polytechnical University, Xi'an 710072, China\\
$^2$State Key Laboratory of Tribology, Department of Mechanical Engineering, Tsinghua University, Beijing 100084, China\\
$^3$Key Laboratory of Advanced Micro-structure Materials (MOE) and\\
School of Physics Science and Engineering, Tongji University, Shanghai 200092, China\\
$^*$Corresponding author: phyfan@nwpu.edu.cn
}

\begin{abstract}Coherent perfect absorber (CPA) was proposed as the time-reversed counterpart to laser: a resonator containing lossy medium instead of gain medium can absorb the coherent optical fields completely. Here, we exploit a monolayer graphene to realize the CPA in a \textbf{\textit{non-resonant}} manner. It is found that quasi-CPA point exists in the terahertz regime for suspending monolayer graphene, and the CPA can be implemented with the assistant of proper phase modulation among two incident beams at the quasi-CPA frequencies. The graphene based CPA is found of broadband angular selectivity: CPA point splits into two frequency bands for the orthogonal $s$ and $p$ polarizations at oblique incidence, and the two bands cover a wide frequency range starting from zero frequency. Furthermore, the coherent absorption can be tuned substantially by varying the gate-controlled Fermi energy. The findings of CPA with non-resonant graphene sheet can be generalized for potential applications in terahertz/infrared detections and signal processing with two-dimensional optoelectronic materials.
\end{abstract}


 ] 

Laser differs from other light sources for that it emits light coherently. The coherence feature of laser light made it unique and essential in modern optoelectronics and photonics. A coherent perfect absorber (CPA), or anti-laser grounded on the time-reversed process of lasing was recently proposed theoretically \cite{1} and demonstrated experimentally \cite{2} in a simple Silicon-resonator. The coherent absorption arises from the coherent modulation on the scattered light beams.  Since the first proposal, relevant coherent modulation assisted processes have attracted considerable research interests with various photonic structures \cite{3,4,5,6,7,8}, e.g., laser absorber and symmetry breaking in ${\mathcal{P T}}$-symmetric optical potentials and strongly scattering systems \cite{3}, unidirectional invisibility in ${\mathcal{P T}}$-symmetric periodic structures \cite{4} and perfect mode (polarization or morphology) conversions \cite{5}. However, the proposed coherent manipulations are all relying on structured resonances \cite{1,2,3,4,5,6,7,8}: either Bragg-type (collective) or Mie-type (local) \cite{9}.

Graphene---carbon atoms in a hexagonal lattice has recently attracted considerable attention for both its fundamental physics and enormous applications. Such a two-dimensional (2D) material is emerging in photonics and optoelectronics \cite{10,11}. As a new optoelectronic material, graphene exhibits much stronger binding of surface plasmon polaritons and supports its relatively longer propagation \cite{12}. Linear dispersion of the 2D Dirac fermions provides ultrawideband tunability through electrostatic field, magnetic field or chemical doping \cite{13,14,15}. Graphene is almost transparent to optical waves \cite{16}, which is one remarkable feature of two-dimensional materials. However, optical insulator \cite{17} similar to gapped graphene \cite{18} for nanoelectronics is frequently required in myriad applications for all-optical systems and components of much miniaturized optical circuits \cite{19}, artificially constructed micro-structure, i.e. metamaterial \cite{20}, as a platform for enhancing light-matter interactions has been employed for this purpose. A recent study reported that optical absorption enhancement can be achieved in periodically doped graphene nanodisks, in which periodic graphene nanodisks are overlying on a substrate or on a dielectric film coating on metal, the excitation of electric mode of the nanodisks together with multi-reflection from the assistants of total internal reflection and metal reflection can result in a complete optical absorption \cite{21}. Graphene micro-ribbons, mantles, nano-crosses, ring resonators and super-lattices have also been attempted for  manipulating the terahertz/infrared waves \cite{22,23,24,25,26,27,28}.

In this paper, we suggest to boost the absorption of terahertz radiations by exploiting the concept of coherent absorption within a \textbf{\textit{non-resonant}} suspending monolayer graphene. It is found that quasi-CPA frequency, which is the necessary formation condition of coherent absorption, does exist in the terahertz regime for monolayer graphene. By introducing a proper two-channel coherent modulation on the input beams, we can perfectly suppress the scattering of coherent beams and thus make a terahertz CPA. The angular selectivity in a wide frequency band and the flexibility of the doping influenced CPA working-frequency of the monolayer graphene sheet are of interests for tunable terahertz/infrared detections and signal modulations. The designed monolayer graphene CPA is shown schematically in Fig. 1, a monolayer graphene is free-standing in vacuum, and it is illuminated by two counter-propagating and coherently modulated input beams ($I_+$ and $I_-$), $O_+$ and $O_-$ are the respective output magnitudes. The monolayer graphene sheet is lying on the $xy$-plane, and throughout our study the graphene sheet is illuminated with $s$-polarized (electric vector $\textit{\textbf{E}}$ is parallel to the $y$-axis) or $p$-polarized (magnetic vector $\textit{\textbf{H}}$ is parallel to the $y$-axis) optical waves, $\theta$ as illustrated in Fig. 1 is the incident angle.

In the monolayer graphene system, the complex scattering coefficients ($O_\pm$) can be related to the two input beams ($I_\pm$) through a scattering matrix, $S_g$, which is defined as:
\begin{equation}
\begin{pmatrix}
O_+ \\
O_-
\end{pmatrix} =S_g
\begin{pmatrix}
I_+ \\
I_-
\end{pmatrix}=
\begin{pmatrix}
t_+ & r_- \\
r_+ & t_-
\end{pmatrix}
\begin{pmatrix}
Ie^{i\phi_+} \\
Ie^{i\phi_-}
\end{pmatrix},
\end{equation}
where $(t/r)_+$ and $(t/r)_-$ are scattering elements of forward (irradiate towards right, $I_+$) and backward (irradiate towards left, $I_-$) beams, since the linear monolayer graphene under investigation is of reciprocity and spatial symmetry, the scattering matrix can be simplified with $t_\pm=t$ and $r_\pm=r$. And in this paper, we only consider phase modulation of the coherent input beams, i.e., the two coherent input beams are set to be of equal amplitude $I$, then the amplitude of the scattering coefficients would be
\begin{equation}
\left|O_+\right|=\left|O_-\right| =\left|tIe^{i\phi_+}+rIe^{i\phi_-}\right|,
\end{equation}

\begin{figure}[t]
\includegraphics[width=8.6cm]{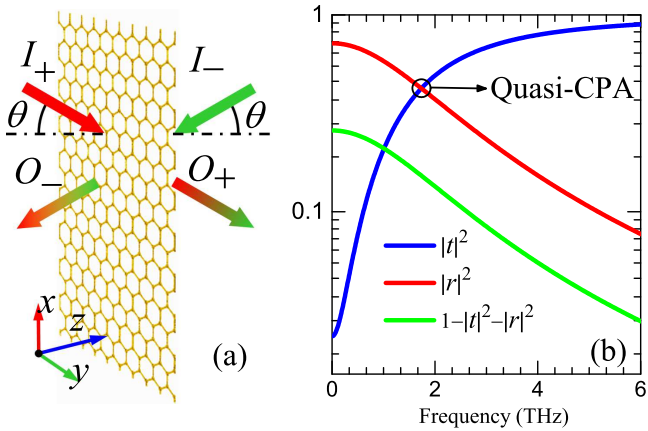}
\caption{\label{fig:epsart} (a) Schematic of a monolayer graphene illustrated by two counter-propagating and coherently modulated input beams ($I_+$ and $I_-$), $O_+$ and $O_-$ are the output intensities. (b) Transmission, reflection and absorption spectra of a doped monolayer graphene ($E_{F}=0.5$ eV) with a Quasi-CPA point at the crossing of transmission and reflection curves.}
\end{figure}

In a terahertz coherent perfect absorber, coherent modulation of the input beams performance is required to inhibit the scatterings and thus stimulate the complete absorption of coherent terahertz beams, which requires $tIe^{i\phi_+}=rIe^{i\phi_-}$, we see that $\left|t\right|=\left|r\right|$ is the necessary condition for acquiring CPA performance.
\begin{figure}[b]
\includegraphics[width=8.6cm]{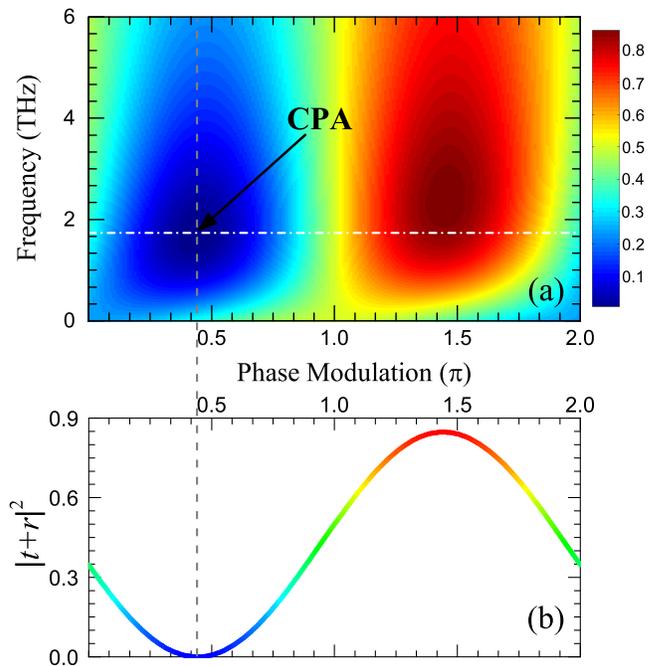}
\caption{\label{fig:epsart} (a) A two-dimensional false-color plot of the normalized total output intensities as a function of frequency and phase modulation, exact CPA point is denoted with a black arrow. (b) Normalized total output intensities as a function of frequency at the CPA frequency.}
\end{figure}

A monolayer graphene can be approximatively treated as an optical interface with complex surface conductivity ($\sigma_g$), since a one-atom-thick graphene sheet is sufficiently thin compared with the concerned wavelength. When the graphene is illuminated, the surface current will be excited by the incident waves, and it is solely determined by the dynamic surface conductivity. Complex coefficients of forward and backward propagating electromagnetic fields through a conductive graphene sheet can be related with the assistance of Ohm's law. The scattering elements $t$ and $r$ at normal incidence are given by
\begin{equation}
t_\bot=\frac{2}{2+\sigma_g \eta_0},
\end{equation}
\begin{equation}
r_\bot=\frac{\sigma_g \eta_0}{2+\sigma_g \eta_0},
\end{equation}
where $\eta_0$ is the wave impedance of free space, and $\sigma_g$, the complex conductivity of the graphene is adopted from a random-phase-approximation (RPA) \cite{29}, which can be well described by a Drude model \cite{30,31} as $\sigma_g=i e^{2}E_{F}/\pi\hbar^{2}\left(\omega+i\tau^{-1}\right)$, especially at heavily doped region and low frequencies (far below Fermi energy), where $E_{F}$ is the Fermi energy away from Dirac point, $\tau=\mu E_{F}/e\upsilon_{F}^2$ is the relaxation rate, in which $\mu=10^{4}$ cm$^2$V$^{-1}$s$^{-1}$ and $\upsilon_{F}\approx10^{6}$m/s are the mobility and Fermi velocity. The transmission, reflection and absorption of a monolayer graphene, obtained from Eqs. (3) and (4) with the aforementioned dynamic conductivity (in our theoretical considerations, we first took $E_{F}=0.5$ eV, as that in Ref. 28, which is quite close to experimental data), are plotted in Fig. 1(b). There exists a frequency ($1.735$ THz), which we call \textit{quasi-CPA} point, where $\left| t\right|^2 = \left| r\right|^2$ implies the formation condition for suppressing the scattering fields to completely absorb coherent input beams of equal-intensity.

To further investigate the coherent absorption process with a monolayer graphene, we plot false-color map of the normalized total output intensity spectra in Fig. 2(a), which shows the detailed dependence on the phase modulation $\Delta \phi=\phi_+ -\phi_-$ of the beams $I_\pm$. It is seen that proper phase modulation ($\Delta \phi=0.435\pi$) of the input coherent beams leads to significant suppression of the scattering outputs at the quasi-CPA frequency ($1.735$ THz). The corresponding normalized total output intensities for the CPA frequency is plotted as a function of phase modulation $\Delta\phi$ in Fig. 2(b). At the quasi-CPA frequency, we can see a substantial modulation of the scattered intensity of the coherent inputs according to the phase difference. The significant reduction of the scattered intensity of about zero for $\Delta \phi=0.435\pi$ implies destructive interference which prevents the coherent beams from escaping the absorbing channel of the monolayer graphene, demonstrating completely absorption of the coherent input beams. A defined convenient figure of merit to measure the modulation in a CPA is the ``modulation depth'' $M(\omega)=\max(O_\pm)/\min(O_\pm)$ \cite{2}. As the results showed in Fig. 2(b), we have $M(1.735$ THz$)\approx 10^6$, which is an impressive modulation owing much to the thin-sheet feature that permits destructive suppression of the fields in the monolayer graphene.
\begin{figure}[t]
\includegraphics[width=8.6cm]{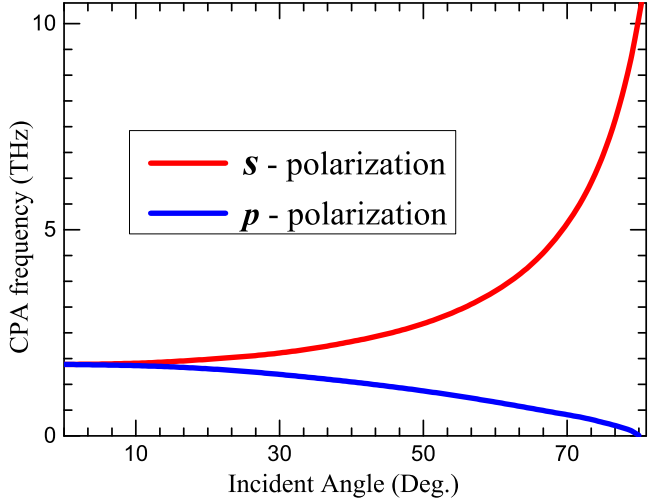}\caption{\label{fig:epsart} CPA frequency dispersion at oblique incidence for $s$ polarization (red solid line) and $p$ polarization (blue solid line).}
\end{figure}

Angular selectivity, or spatial dispersion, is a fundamental and important property for the realization of light selection \cite{32}. We found that the monolayer graphene based CPA is of angularly sensitivity, which can be beneficial for broadband angular tunability. For oblique incidence, both orthogonal $s-$ and $p-$ polarized modes should be considered for the scattering problem. We get the quasi-CPA frequencies by employing the transfer matrix formalism, figure 3 clearly shows that the CPA frequencies split into two frequency branches for orthogonal $s$ and $p$ polarizations: for $s$-polarization, the CPA has a blue shift with respect to the normal incident CPA point, forming the upper band; while for the $p$-polarization, the CPA has a red shifts with respect to the normal incident CPA point, forming the lower band. The two bands touch each other and cover a wide band starting from zero frequency.
\begin{figure}[t]
\includegraphics[width=8.3cm]{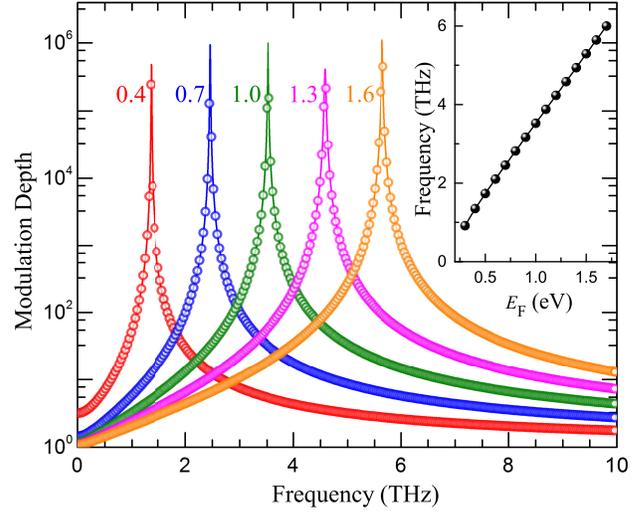}
\caption{\label{fig:epsart} Tunability of the monolayer graphene CPA: Spectra of modulation depth for different doping levels ($E_{F}=0.4$ eV,$0.7$ eV,$1.0$ eV,$1.3$ eV and $1.6$ eV). The inset shows doping dependent of the CPA frequency.}
\end{figure}

The dependence of Fermi energy on the charge-carrier density $n$ is predicted (with a quantum theoretical treatment of electron dynamic response in graphene) as: $E_{F}=\hbar \upsilon_{F}\sqrt{\pi\left|n\right|}$, which has been tested experimentally by direct terahertz to mid-IR spectroscopic measurements. The charge-carrier density can be easily changed through electrostatic doping, which makes graphene promising for wide-tunable and broadband optoelectronic and photonic applications. Figure 4 displays the calculated ``modulation depth'' spectra of the monolayer graphene for different doping levels or Feimi energies. A sharp peak at $1.356$ THz is seen from the modulation depth spectrum for $E_F=0.4$ eV, corresponding to the quasi-CPA frequency where scattering can be significantly suppressed, demonstrating the maximum ``modulation depth'' on the curve. With the increasing of the electrostatic doping we get higher charge-carrier concentration and thus higher Fermi energies, and we see from Fig. 4 a blue shift of the peak frequency, or the quasi-CPA frequency, in the ``modulation depth'' spectra. The shift processing can be understood as follows: graphene will have a higher Drude weight by increasing the charge-carrier concentration of graphene Dirac fermion, and the graphene with reenforced metallicity will be even more scattering, increasing the reflection and decreasing the transmission, in that way quai-CPA point where $\left|t\right|=\left|r\right|$ will shift to higher frequencies. As showed in the inset of Fig. 4, the quasi-CPA frequency can be tuned from $0.916$ THz to $5.998$ THz when the doping level varies from $0.3$ eV to $1.7$ eV.

In summary, we show that non-resonant two-dimensional carbon material---graphene can be employed for perfectly suppressing scattering of terahertz radiations for coherent perfect absorber (CPA). Meanwhile the CPA with a monolayer graphene is of broadband angular selectivity, and two frequency branches for orthogonal $s$ and $p$ polarizations degenerate at the quasi-CPA frequency for normal incidence and combine a broad band starting from zero frequency. Furthermore, the CPA can be tuned in a wide range via changing the charge-carrier density through electrostatic doping (gate voltage). The idea of graphene based CPA is general for two-dimensional conductive material systems, the non-resonant CPA can be extended to mid-infrared or even near-infrared regime if we can achieve two-dimensional materials with higher charge-carrier density, and we expect potential applications of coherent modulations in terahertz/infrared detections and signal processing with two-dimensional materials.

The authors would like to acknowledge financial support from the NSFC (Grants No. 11174221, 11372248, 61101044 and 61275176), and the National 863 Program of China (Grant No. 2012AA030403).



\begin{thebibliography}{99}


\bibitem{1} Y. D. Chong, L. Ge, H. Cao, and A. D. Stone, Phys. Rev. Lett. \textbf{105}, 053901 (2010).
\bibitem{2} W. Wan, Y. Chong, L. Ge, H. Noh, A. D. Stone, and H. Cao, Science \textbf{331}, 889 (2011).
\bibitem{3} S. Longhi, Phys. Rev. A \textbf{82}, 031801 (2010).
\bibitem{4} Z. Lin, H. Ramezani, T. Eichelkraut, T. Kottos, H. Cao, and D. N. Christodoulides, Phys. Rev. Lett. \textbf{106}, 213901 (2011).
\bibitem{5} H. Noh, Y. Chong, A. D. Stone, and H. Cao, Phys. Rev. Lett. \textbf{108}, 186805 (2012); M. Crescimanno, N. J. Dawson, and J. H. Andrews, Phys. Rev. A \textbf{86}, 031807 (2012).
\bibitem{6} J. Zhang, K. F. MacDonald, and N. I. Zheludev, Light: Science \& Applications \textbf{1}, e18 (2012); J. Hao, L. Zhou, and M. Qiu, Phys. Rev. B \textbf{83}, 165107 (2011); F. Liu Y. D. Chong, S. Adam, and M. Polini, arXiv:1402.2368.
\bibitem{7} M. Pu, Q. Feng, C. Hu, and X. Luo, Plasmonics \textbf{7}, 733 (2012).
\bibitem{8} Y. Sun, W. Tan, H. Li, J. Li, and H. Chen, Phys. Rev. Lett. \textbf{112}, 143903 (2014).
\bibitem{9} C. M. Soukoulis and M. Wegener, Science \textbf{330}, 1633 (2010).
\bibitem{10} F. Bonaccorso, Z. Sun, T. Hasan, and A. C. Ferrari, Nat. Photonics \textbf{4}, 611 (2010).
\bibitem{11} H. Yan, X. Li, B. Chandra, G. Tulevski, Y. Wu, M. Freitag, W. Zhu, P. Avouris, and F. Xia, Nat. Nanotechnol. \textbf{7}, 330 (2012).
\bibitem{12} M. Jablan, H. Buljan,and M. Solja\v{c}i\'{c}, Phys. Rev. B \textbf{80}, 245435 (2009).
\bibitem{13} T. Low and P. Avouris, ACS Nano \textbf{8}, 1086 (2014).
\bibitem{14} A. Vakil and N. Engheta, Science \textbf{332}, 1291 (2011).
\bibitem{15} F. H. L. Koppens, D. E. Chang, and F. J. Garc\'{i}a de Abajo, Nano Lett. \textbf{11}, 3370 (2011).
\bibitem{16} R. R. Nair, P. Blake, A. N. Grigorenko, K. S. Novoselov, T. J. Booth, T. Stauber, N. M. R. Peres, and A. K. Geim, Science \textbf{320}, 1308 (2008).
\bibitem{17} E. Yablonovitch, Sci. Am. \textbf{285}, 46 (2001).
\bibitem{18} Z. Zhang, H. Li, Z. Gong, Y. Fan, T. Zhang, and H. Chen, Appl. Phys. Lett. \textbf{101}, 252104 (2012).
\bibitem{19} N. Engheta, Science \textbf{317}, 1698 (2007); Y. Fan, J. Han, Z. Wei, C. Wu, Y. Cao, X. Yu, and H. Li, Appl. Phys. Lett. \textbf{98}, 151903 (2011).
\bibitem{20} D. R. Smith, J. B. Pendry, and M. C. K. Wiltshire, Science \textbf{305}, 788 (2004).
\bibitem{21} S. Thongrattanasiri, F. H. L. Koppens, and F. J. Garc\'{i}a de Abajo, Phys. Rev. Lett. \textbf{108}, 047401 (2012).
\bibitem{22} R. Alaee, M. Farhat, C. Rockstuhl, and F. Lederer, Opt. Express \textbf{20}, 28017 (2012).
\bibitem{23} S. He, X. Zhang, and Y. He, Opt. Express \textbf{21}, 30664 (2013).
\bibitem{24} H. Yan, T. Low, W. Zhu, Y. Wu, M. Freitag, X. Li, F. Guinea, P. Avouris, F. Xia, Nat. Photonics \textbf{7}, 394 (2013).
\bibitem{25} P. Y. Chen and A. Al\`{u}, ACS Nano \textbf{5,} 5855 (2011).
\bibitem{26} H. Cheng, S. Chen, P. Yu, J. Li, L. Deng, and J. Tian, Opt. Lett. \textbf{38}, 1567 (2013).
\bibitem{27} P. Liu, W. Cai, L. Wang, X. Zhang, and J. Xu, Appl. Phys. Lett. \textbf{100}, 153111 (2012); W. Wang, J. Phys.: Condens. Matter \textbf{24}, 402202 (2012); N. Papasimakis, S. Thongrattanasiri, N. I Zheludev and F. J. Garc\'{i}a de Abajo, Light: Science \& Applications \textbf{2}, e78 (2013); Y. Fan, Z. Wei, Z. Zhang, and H. Li, Opt. Lett. \textbf{38}, 5410 (2013).
\bibitem{28} Y. Fan, Z. Wei, H. Li, H. Chen, and C. M. Soukoulis, Phys. Rev. B \textbf{88}, 241403(R) (2013).
\bibitem{29} V. P. Gusynin, S. G. Sharapov and J. P. Carbotte, J. Phys.: Condens. Matter \textbf{19}, 026222 (2007).
\bibitem{30} Z. Q. Li, E. A. Henriksen, Z. Jiang, Z. Hao, M. C. Martin, P. Kim, H. L. Stormer, and D. N. Basov, Nat. Phys. \textbf{4}, 532 (2008).
\bibitem{31} J. Horng, C.F. Chen, B. Geng, C. Girit, Y. Zhang, Z. Hao, H. A. Bechtel, M. Martin, A. Zettl, M. F. Crommie, Y. R. Shen, and F. Wang, Phys. Rev. B \textbf{83}, 165113 (2011).
\bibitem{32} Y. Shen, D. Ye, I. Celanovic, S. G. Johnson, J. D. Joannopoulos, and M. Solja\v{c}i\'{c}, Science \textbf{343}, 1499 (2014).

\end{thebibliography}
\end{document}